\documentclass[reprint, amsmath, amssymb, aps, superscriptaddress, prb]{revtex4-2}
\usepackage{graphicx}
\usepackage{lipsum}

\usepackage{xcolor}

\begin{document}

\begin{abstract}
An experimental study of Landau levels (LLs) in a system of two-dimensional massless Dirac fermions based on a critical thickness HgTe quantum well has been carried out. The magnetotransport and the capacitive response have been investigated simultaneously. It is shown that the formation of Shubnikov-de Haas (SdH) oscillations associated with odd $\nu$ filling factors occurs in a magnetic field whose strength grows monotonically with $\nu$. This behavior is consistent with calculations of the electron spectrum, which predicts a decrease in cyclotron gaps with increasing $\nu$.
Oscillations with even filling factors, corresponding to spin gaps, behave less trivially. First, the SdH oscillations with filling factors of 4 and higher are resolved in a magnetic field that is 2-2.5 times smaller than the field required to resolve neighboring SdH oscillations with odd filling factors of 3 and higher. This indicates a significant increase in the size of the spin gap caused by an interface inversion asymmetry (IIA) leading to Dirac cone splitting in a zero magnetic field. Using the spin splitting value $\gamma$ as a fitting parameter, we obtained the best agreement between experimental data and calculations at $\gamma=1.5$\,meV. Next, spin splitting for the zeroth and first LLs is observed in 2-3 times stronger magnetic fields than for the other levels, indicating an  increase in disorder near the Dirac point, due to the lack of screening.
\end{abstract}

\title{Spin Splitting and Disorder in HgTe-Based Massless Dirac Fermion Landau Levels}

\author{D.\,A.~Kozlov}
\affiliation{Experimental and Applied Physics, University of Regensburg, D-93040 Regensburg, Germany}
\affiliation{Rzhanov Institute of Semiconductor Physics, 630090 Novosibirsk,
	Russia}

\author{J.Ziegler}
\affiliation{Experimental and Applied Physics, University of Regensburg, D-93040 Regensburg, Germany}

\author{N.\,N.\,Mikhailov}
\affiliation{Rzhanov Institute of Semiconductor Physics, 630090 Novosibirsk,
	Russia}


\author{Z.\,D.\,Kvon}
\affiliation{Rzhanov Institute of Semiconductor Physics, 630090 Novosibirsk,
	Russia}
\affiliation{Novosibirsk State University, 630090 Novosibirsk, Russia}

\author{D.\,Weiss}
\affiliation{Experimental and Applied Physics, University of Regensburg, D-93040 Regensburg, Germany}

\maketitle

Massless Dirac fermion systems, similar to graphene, can be realized in 
HgTe quantum wells with a critical thickness of 6.3$\ldots$6.6\,nm. At this thickness the energy spectrum changes from direct to inverted \cite{Buttner2011}.  
Such systems are of interest because of their linear dispersion law and their strong spin-orbit interaction. In this system classical transport \cite{Tkachov2011_DFMoblitiy, Kozlov2012_WeakLoc, Dobretsova2016, Dobretsova2018, Gusev2022ChannelNetwork}, quantum transport \cite{Buttner2011, Kozlov2012_WeakLoc, Pakmehr2014, Kozlov2014QHE77K, Kozlov2014QHELowField, Khouri2016, Gusev2017, Gusev2022QHE}, and cyclotron resonance  \cite{Kvon2012CR, Olbrich2013CRinPC, Dziom2022, Zoth2014CRinPC, Ludwig2014CR} were investigated, the density of states (DoS) was measured \cite{Kozlov2016DFCapacitance} and the quantization of the  Faraday rotation was discovered \cite{Shuvaev2016FaradayQuantum}.

Although numerous methods have been applied and diverse results have been obtained, with a considerable body of work dedicated to band structure and Landau level (LL) calculations \cite{Raichev2012, Scharf2012, Bernevig2006Hamiltionian, Buttner2011, Novik2005Hamiltionian, Tarasenko2015, Durnev2016, Dziom2022, Olbrich2013CRinPC}, the experimental data currently available on the characteristics of these levels remains remarkably incomplete. On the one hand, the study of cyclotron resonance has allowed to confirm the existence of the Dirac spectrum and the non-equidistance of LLs \cite{Kvon2012CR, Olbrich2013CRinPC, Zoth2014CRinPC}. On the other hand, the cyclotron resonance obeys selection rules prohibiting spin flips when the orbital number changes by one. It is, therefore,  insensitive to Zeeman splitting. Magnetotransport measurements are not subject to these limitations. However, previous research has primarily focused on the quantum Hall effect (QHE) in high magnetic fields \cite{Kozlov2014QHE77K, Khouri2016} or on the hole side, where Fermi level pinning and the observation of ultra-long QHE plateaus for light holes are observed due to the presence of side valleys with heavy holes  \cite{Kozlov2014QHELowField, Gusev2022QHE}. At the same time, the region of weak magnetic fields, where the features of the Dirac spectrum should be most evident, is virtually unexplored experimentally.

The charge neutrality point is of particular interest. In quantizing magnetic fields, the zero LL is a characteristic fingerprint of Dirac fermion system \cite{Zhang2005, Ponomarenko2010}. In graphene, due to the presence of two valleys and spin, it is 4-fold degenerate, but because of the interaction effects the degeneracy can be lifted \cite{Yu2013}. However, due to the small magnitude of the interaction effects and the small value of the g-factor, the degeneracy lifting is observed at rather high magnetic fields.
In the QHE regime, the conductivity of graphene at the Dirac point  
can be due to both the weakly conducting bulk and the counterpropagating dissipative edge states \cite{Abanin2007}. A different behavior can be expected in HgTe Dirac fermion quantum wells because of the large Zeeman splitting which could open a gap at the Dirac point.
The QHE near the charge neutrality point  has been studied in detail in semi-metallic HgTe QWs \cite{Gusev2010, Raichev2012-2} and, to a lesser extent, in QWs of critical thickness \cite{Gusev2017, Gusev2022-2} or close to it \cite{Ma2015}, but with a focus on strong magnetic fields or mesoscopic samples. Thus, the behavior of the system of massless Dirac fermions at the charge neutrality point at the transition from weak to quantizing magnetic fields remains poorly understood. 

The present work focuses on the study of Landau levels in a (013)-oriented 6.6\,nm thick HgTe quantum well (see Fig.~1(a)) in quantizing but relatively weak (less than 3\,Tesla) magnetic fields at temperatures of 1.5-10\,K. We performed combined magnetotransport and capacitance measurements. Due to the high sensitivity of the transport measurements and the possibility to measure the DoS directly by the capacitive technique, we were able to study the zero LL in detail and to demonstrate its splitting in a strong field. The samples investigated were 10-pin gated Hallbars with a size of 450$\times$50\,$\mu$m and a total capacitance of 36\,pF. Magnetotransport measurements were performed using a 4-terminal scheme with lock-in detection at a frequency of 4-12\,Hz and a drive current of 10-100\,nA, which prevents heating effects. The capacitive measurements were performed according to a scheme similar to that described in \cite{Kozlov2016DFCapacitance, Kozlov2016Capacitance3DTI}: a small oscillating voltage $V_{\rm ac}$ was applied to the gate at frequency $f$ against a constant bias $V_g$, while the quantum well was at zero potential. The magnitude of the AC current, phase-shifted by 90 degrees with respect to the AC voltage, reflected the capacitance of the structure. The parameters $V_{\rm ac}$, $f$ and $T$ were varied as a function of the magnetic field in order to eliminate both resistive effects and the effect of DoS smearing by  temperature and drive voltage, while achieving the highest possible signal-to-noise ratio. The parameters used were $V_{\rm ac} = 20\ldots100$\,mV, $f = 83\ldots4$\,Hz, $T = 1.5\ldots10$\,K, where the first number of the respective range corresponds to  zero magnetic field and the second one  to $B = 3$\,T.

\begin{figure}
	\centering
		\includegraphics[width=0.9\columnwidth]{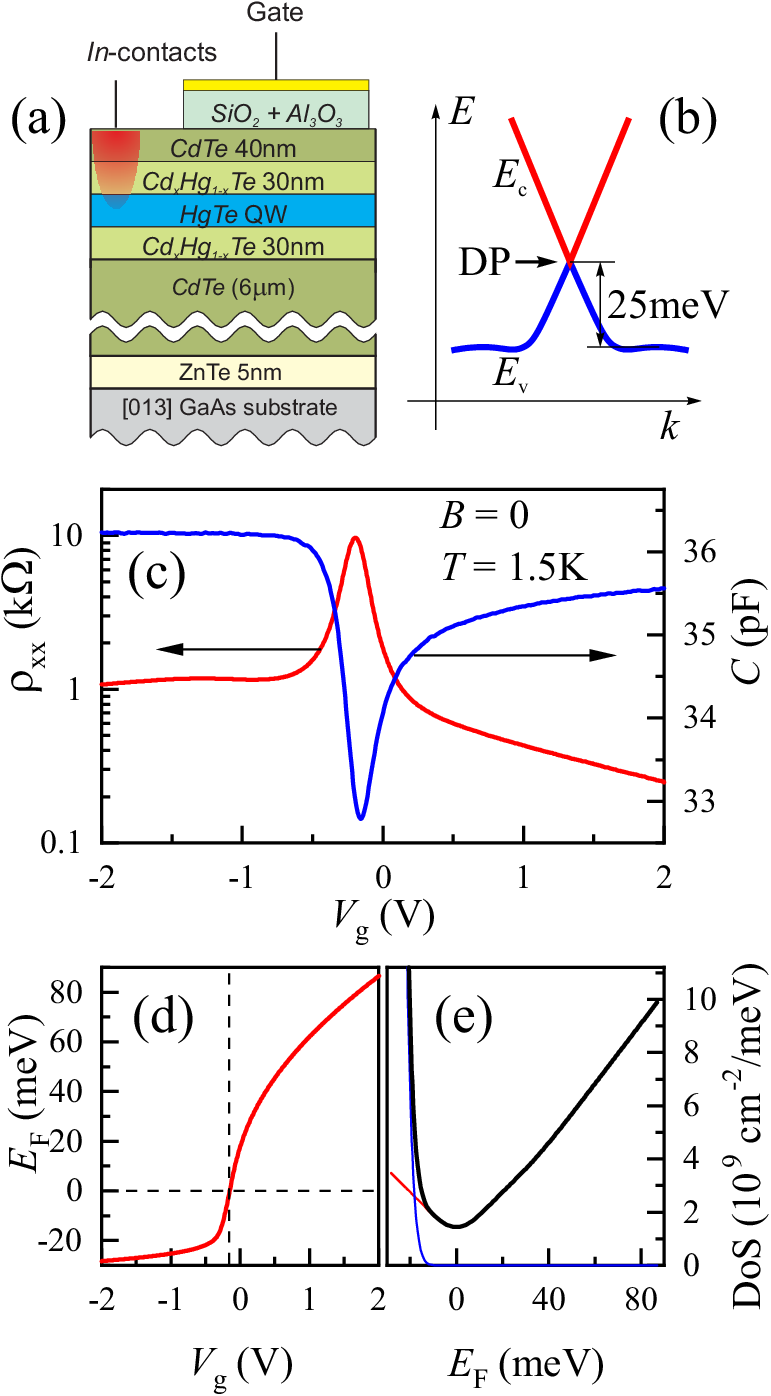}
	\caption{(a)~ Cross section of the investigated structure. (b)~ Schematic band structure of the system. The side valleys in the valence band are located 
    $\approx25$\,meV below the Dirac point. (c)~Gate voltage dependence of the longitudinal resistance $\rho_\text{xx}$ (red line, left axis) and the capacitance (blue line, right axis) at $B = 
    0$  and $T = 1.5\,$K. (d) and (e)~Fermi energy vs. gate voltage (d) and the DoS vs. Fermi energy (e) extracted from the capacitance data.}
	\label{fig:Fig1}
\end{figure}

The measured dependencies of $\rho_{\rm xx}(V_g)$ and $C(V_g)$ are shown in Fig.~1(c). The Dirac point (DP) is located at $V_g=-0.18$\,V near the maximum of $\rho_{\rm xx}$ and the minimum of $C$. As one moves away from the DP to the right, the electron density in the system increases and a smooth decrease in resistance and increase in capacitance is observed. From the measured capacitance and following the previously developed approach \cite{Kozlov2016DFCapacitance}, we extracted the dependencies of the Fermi energy $E_{\rm F}$ (Fig.~1(d)) and the DoS (Fig.~1(e)) on the gate voltage. On the electron side, we can clearly see that the DoS depends almost linearly on the Fermi energy. It reaches a value of 80\,meV at $V_g = 2$\,V. On the left side of the DP, Dirac holes coexist with heavy holes \cite{Kozlov2014QHELowField, Gusev2022QHE, Kozlov2016DFCapacitance}. The presence of heavy holes leads to a rapid increase of the DoS, and thus to the pinning of the Fermi level at about $-25$\,meV below the DP (Fig.~1(d)) and causes the measured capacitance to saturate to the value of the geometric capacitance (Fig.~1(c)).

\begin{figure}
	\centering
	\includegraphics[width=0.95\columnwidth]{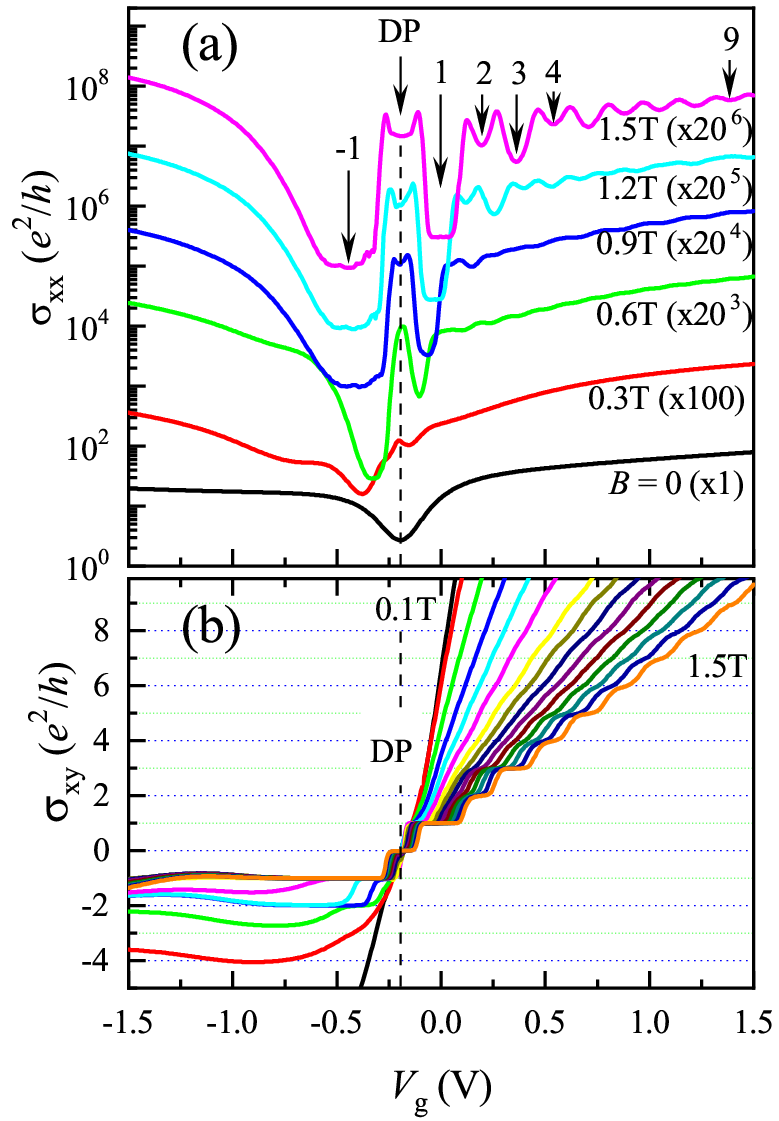}
	\caption{(a)~ Gate voltage dependence of the longitudinal conductivity $\sigma_{\rm xx}$, measured for $B = 0$, $0.3$ $\ldots1.5$\,T and $T = 1.5$\,K. For clarity,  curves measured at non-zero magnetic fields are multiplied by the factor shown in parentheses. The minima and corresponding filling factors are indicated by the numbered vertical arrows. (b)~ Gate voltage dependence of the Hall conductivity $\sigma_{\rm xy}$, measured for $B = 0.1\ldots1.5$\,T in steps of $0.1$\,T. In both panels "DP" denotes the Dirac point.}
	\label{fig:Fig2}
\end{figure}

Figure 2 shows the dependencies of the conductivity tensor components $\sigma_{\rm xx}(V_g)$ and $\sigma_{\rm xy}(V_g)$  in magnetic fields up to 1.5\,T, calculated by tensor inversion from the measured $\rho_{xx}$ and $\rho_{xy}$ data. These dependencies show a transition from classical transport to the QHE regime, accompanied by the formation of a series of distinct minima in $\sigma_{\rm xx}$ and plateaus in $\sigma_{\rm xy}$. The most striking effect is the strong asymmetry between electrons and holes: first, the QHE on the hole side is formed in a magnetic field of only 0.4\,T (and even at 0.15\,T when measured at lower temperatures \cite{Kozlov2014QHELowField}), while on the electron side it requires twice the field. Second, the plateaus on the hole side are anomalously long, and 
from 0.7 to 1.5\,T only one plateau with fixed $\sigma_{\rm xy}=-e^2/h$ is observed. Both features are explained by the coexistence of light and heavy holes \cite{Kozlov2014QHELowField, Gusev2022QHE, Yakunin2020}. The anomalous length of the plateau is related to the exchange light and heavy holes,
implementing the QHE reservoir model. A similar effect was later discovered in graphene on some substrates and is known as "giant QHE" \cite{Kudrynskyi2017GiantQHEGraphene}. The second feature, i.e., the formation of the plateau at ultra-low fields is associated with an effective screening of the random potential by heavy holes leading to a significant increase of the quantum lifetime.

The QHE on the electron side in Fig.~2(b) shows up to five electronic plateaus of $\sigma_{\rm xy}$, which become blurred for higher filling factors. This behavior reflects the peculiarity of Dirac fermion systems, where the distance between neighboring LLs decreases with increasing $\nu$. It could be explained by the magnetic field dependence of the LLs in the simplest approximation, fitted to the the 4-zone Bernevig-Hughes-Zhang (BHZ) Hamiltonian \cite{Bernevig2006Hamiltionian, Buttner2011}
\begin{equation}
	E_n^\pm (B) = \alpha \sqrt{n B} \pm g \mu_B B /2,
	\label{spectrum}
\end{equation}
where $n = 0, 1, 2\ldots$ is the Landau quantum number, $\alpha = 25$\,meV$\cdot$T$^{-1/2}$ is a numerical coefficient, $g = 50$ is the effective g-factor, $\mu_B$ is the Bohr magneton, with $g \mu_B = 3.5$\,meV$\cdot$T$^{-1}$, and $\pm$ denotes different spin orientations. From formula (1) it can be seen that in weak magnetic fields the distance between Landau gaps characterized by odd filling factors is larger than the distance between spin-split gaps (even filling factor) with the maximum distance for filling factor $\nu=1$. These calculations agree with the experimental  $\sigma_{\rm xx}(V_g)$ values at $B = 1.5$\,T (Fig.~2(a)): the deepest minima of the conductances are observed at $\nu=1$ and $\nu=3$, while the minima at $\nu=2, 4$, and 6 are significantly less deep. At higher filling factors, the amplitude of the conductivity oscillations associated with even and odd filling factors become equal. For the most accurate calculation it is necessary to use the more complicated 6- or even 8-band Kane Hamiltonian \cite{Novik2005Hamiltionian, Raichev2012}. However, on the electron side and in weak magnetic fields (below 2\,T) very similar results are obtained with all three approaches.

\begin{figure}
	\centering
	\includegraphics[width=0.95\columnwidth]{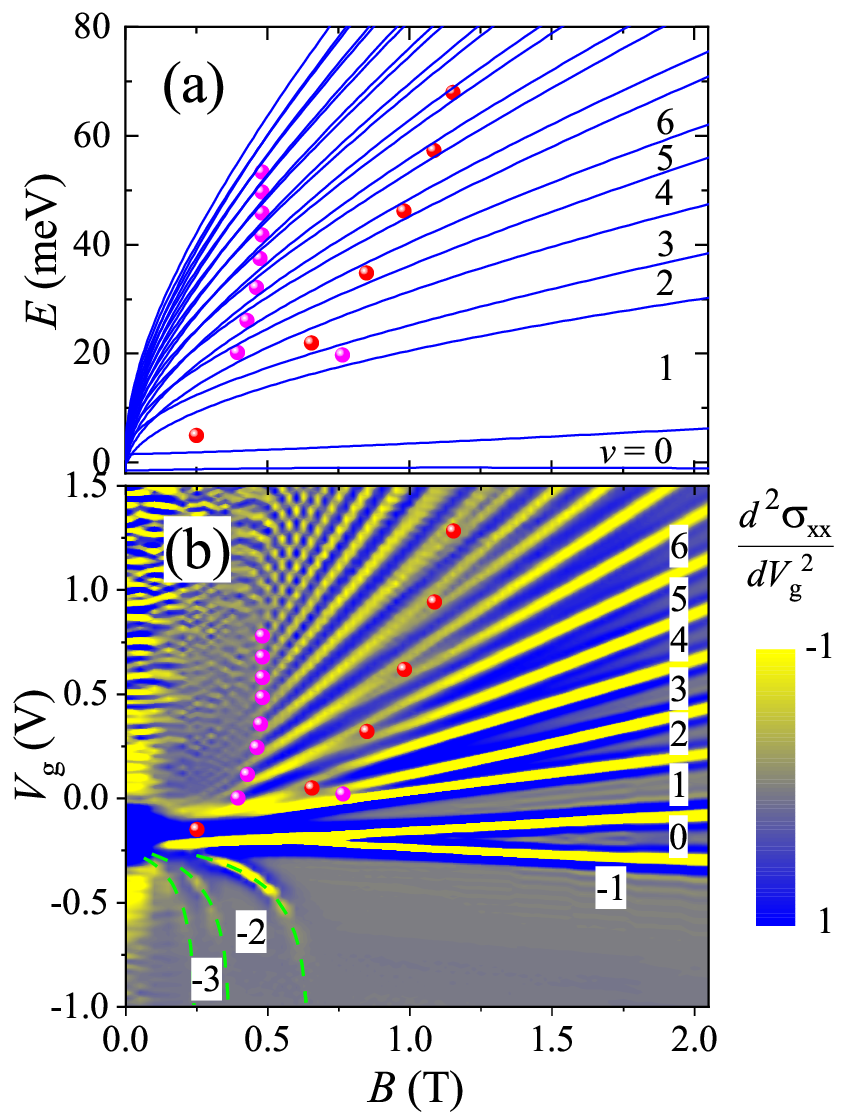}
	\caption{(a)~LLs calculated on the basis of the BHZ model with interface inversion asymmetry taken into account with cone splitting $\gamma = 1.5$\,meV \cite{Tarasenko2015,Durnev2016}. Numbers denote filling factors, counting the number of occupied Landau levels. The  magnetic field axis is the same as in panel (b). (b)~Two-dimensional map of the second derivative $d^2\sigma_{\rm xx}(B,V_g)/dV_g^2$. The blue color corresponds to minima of $\sigma_{\rm xx}$, yellow to maxima. Numbers, as in panel (a), denote the corresponding filling factors. Green dashed lines mark the behavior of Dirac holes LLs \cite{Gusev2022QHE}. In  both panels the dots indicate the points $B_{\nu }$ where the minima of the conductivity oscillations with the filling factor $\nu$ appear first: red dots correspond to odd filling factors, magenta dots correspond to even ones. The position $B_{\rm 1}$ of the first red dot was determined from the $\sigma_{\rm xx}(V_g)$ data in Fig.\,\ref{fig:Fig2}.}
	\label{fig:Fig3}
\end{figure}

Let us analyze the behavior of the measured conductivity oscillations in the limit of weak magnetic fields. Fig.~3(b) shows a two-dimensional map of the second derivative of the conductance with respect to the gate voltage $d^2\sigma_{\rm xx}(B,V_g)/dV_g^2$. From the depth of the minima at different filling factors, one can estimate the size of the corresponding energy gap. The red and  magenta dots in Fig.~3 indicate the points $B_{\nu }$ where minima of the conductivity with filling factor  $\nu$ first occur. For odd filling factors (red dots), a clear trend of $B_{\nu}$ toward increasing $B$ with increasing $\nu$ is observed. This is consistent with formula (1) and reflects the decreasing energy gap for larger $\nu$. The behavior of $B_{\nu}$ for even filling factors (magenta dots), reflecting spin gaps, turns out to be less trivial. For $\nu=2$ the gap opens at a field of $0.8$\,T, which exceeds both $B_{\rm 1} = 0.25$\,T and $B_{\rm 3} = 0.65$\,T and thus qualitatively agrees with the theoretical expectation described by Eq.\,\ref{spectrum}  However, all subsequent spin gaps open in a magnetic field smaller than $0.5$\,T, i.e. 2-2.5 times smaller than the field for the neighboring odd filling factors. 

In order to explain the observed behavior of the spin gaps, we performed the LL calculations based on the 4-zone BHZ Hamiltonian with an additional factor taken into account, namely the $B = 0$ Dirac cone splitting. The splitting naturally comes from the IIA \cite{Tarasenko2015,Durnev2016} and its magnitude $\gamma$ was used as a fitting parameter. The best fit was obtained for $\gamma = 1.5$\,meV and the calculations result is shown in Fig.~3(a). Note, that the optimal value of $\gamma$ varies from 1.3 to 1.7\,meV for different energy ranges, giving an average value of 1.5\,meV. When IIA is taken into account, the calculation agrees well with the experimental data. This is illustrated by 
the red and magenta dots in Fig.~3(a) which are taken from panel (b). 
It can be seen that the distance between neighboring LLs for odd (red) and even (magenta) filling factors are quite similar, if compared at same Fermi energy (i.e gap for $\nu=5$ should be compared with $\nu=8$). The comparison at the same energy is essential, because the efficiency of impurities screening and thus the LLs' broadening depend on energy. The magnitude of disorder has its maximum near the DP and can reach 10-15\,meV \cite{Kozlov2016DFCapacitance}.

To check the last assumption we studied the zero LL. The nature of the QHE state with $\nu=0$ differs from all other QHE states with integer filling factors. In Dirac fermion systems, the $\nu=0$ state may be formed either by a spin-degenerate half-occupied zero LL or its degeneracy could be lifted because of the spin splitting \cite{Abanin2007}. However, in both cases the values of $\sigma_{\rm xx}$ and $\sigma_{\rm xy}$  tend to be zero due to bulk localization. Alternatively, the spin splitting of the zero LL could be accompanied by the formation of two counterpropagating (electron and hole) edge channels, which could
 enhance the conductivity. However, these channels are not protected from backscattering and therefore the conductivity remains low. Note, that the double peak structure of $\sigma_{\rm xx}$ shown in Fig.~2(a) stems from resistivity tensor inversion and is not a signature of spin splitting. Thus, the local transport response turns out to be weakly sensitive to the Landau zero level splitting. To solve this problem, we used capacitive magnetospectroscopy, which directly probes the value of the DoS for arbitrary filling factors.

\begin{figure}
	\centering
		\includegraphics[width=0.95\columnwidth]{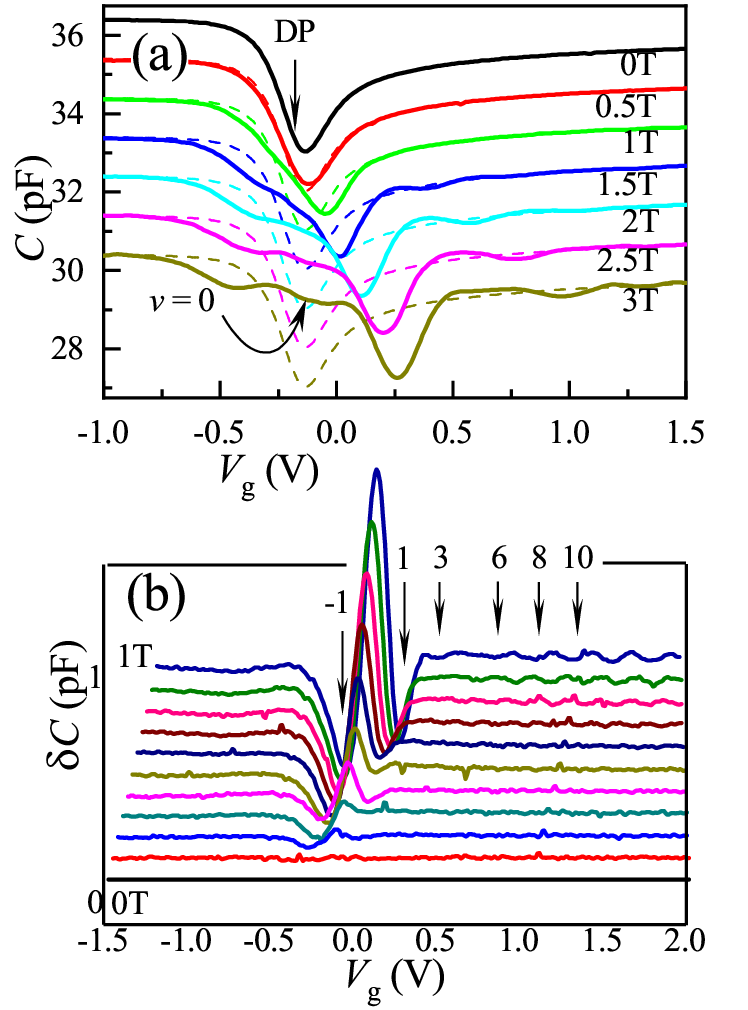}
	\caption{(a)~ Gate voltage dependence of the capacitance $C$, measured at $B = 0$, $0.5$ $\ldots3$\,T and $T = 1.5-10$\,K. For clarity, the curves 
		are shifted vertically. Dashed lines show the zero-field traces as a guide. The Dirac point "DP" is marked by a vertical arrow , "$\nu = 0$" marks the local capacitance minima, associated with the  zeroth Landau level splitting at $B = 3$\,T. (b)~The differential capacitance $\delta C(V_g, B) = C(V_g, B) - C(V_g,B=0)$ measured at $B = 0$, $0.1$ $\ldots1$\,T and $T = 1.5$\,K. The curves for finite $B$ are shifted both on the $x$ and $y$ axes to enhance visibility. The numbers denote the filling factors $\nu$.}
	\label{fig:Fig4}
\end{figure}

The results of the magnetocapacitance measurements $C(V_g)$ are shown in Fig.~4(a). The pronounced SdH oscillations in $C(V_g)$ can only be observed from $B = 1.5$\,T on due to the lower sensitivity compared to transport measurements. To improve the signal, the calculated differential signal $\delta C = C(V_g,B) - C(V_g, B=0)$ is shown in Fig.~4(b). In the differential signal, the SdH oscillations start to appear around 0.8\,T. However, even at $B = 1$\,T not all gaps are resolved: only the largest Landau gaps with corresponding filling factors $\nu$ of 1 and 3 are observed, while the oscillations at higher gate voltages reflect the spin gaps with filling factors 6, 8, 10 and so on. The observed behavior is consistent with the magnetotransport data and the LL calculations. The absence of some minima can be explained by their smaller energy gaps. 

The zeroth LL, a characteristic  and the most remarkable footprint of Dirac fermions, appears in the capacitance signal already at $B=0.2$\,T and is the dominant maximum in Fig.~4(b).  
However, at fields below 1\,T there is not even a hint of Zeeman splitting. The zeroth LL becomes clearly visible as a separate maximum in the non-differential capacitance signal only from $B=2$\,T on. 
At this field all SdH oscillation minima except $\nu=0$ and 2 are already resolved. Finally, the splitting of the zeroth LL appears at $B \sim 3$\,T, where a small and broad minimum can be observed close to the Dirac point. That the spin splitting of the zeroth LLs occurs at a significantly larger magnetic field than for any other one, supports the hypothesis of stronger disorder and broadening. According to our calculations, the value of the spin gap for $\nu = 0$ at $B = 3$\,T is around 10\,meV, which is 2 times bigger then the gap for $\nu = 1$, clearly seen in the capacitance at $B=1$\,T. 
Note that the small deviation of the QW thickness from the critical value could open a small gap between valence and conduction bands also affecting to the splitting of zero LL, however our previous measurements at low temperatures up to 0.2\,K \cite{Kozlov2012_WeakLoc,Kozlov2014QHELowField} and zero magnetic field proved that this energy gap is absent or insignificant.

In summary, the experiments show that for odd filling factors $\nu$ the SdH oscillations are resolved at magnetic field values that increase monotonically with increasing  $\nu$.
This  is consistent with the simplest model (Eq.\,1) describing LLs in Dirac fermion systems. The behavior of SdH oscillations with even filling factors, i.e. those associated with spin gaps, differs significantly from the simplest model. 
First, SdH oscillations with even filling factors of 6 and higher are formed in a magnetic field that is 2-2.5 times smaller than the field required to form neighboring SdH oscillations with odd filling factors.
Second, the oscillations for small even $\nu$ appear at much stronger field, reaching its maximum for $\nu=0$ at a magnetic field of $\sim$\,3\,T. The observed spin gaps are explained by the presence of an interface inversion asymmetry \cite{Tarasenko2015,Durnev2016} with a magnitude of $\gamma = 1.5$\,meV and enhanced disorder at the Dirac point. The obtained value of $\gamma$ is almost an order of magnitude smaller than expected from an interfacial atomistic  calculation \cite{Tarasenko2015}, but it qualitatively agrees with a recent THz spectrocopy study \cite{Dziom2022} with an even smaller value of 0.6\,meV.  

We are grateful to E. L. Golub for useful discussions. The work was supported by the European
Research Council (ERC) under the European Union’s Horizon
2020 research and innovation program (Grant Agreement No.
787515, “ProMotion”).

\bibliography{references}

\end{document}